\documentclass[a4paper,11pt]{article}
\pdfoutput=1 

\usepackage{jheppub} 

\usepackage[T1]{fontenc} 
\usepackage[utf8]{inputenc}
\usepackage{textcomp}
\usepackage{graphicx}
\usepackage{amsmath}
\usepackage[version=4]{mhchem}
\usepackage{siunitx}
\usepackage{longtable,tabularx}
\usepackage[english]{babel}
\usepackage{amsmath}
\usepackage{graphicx}
\usepackage{amssymb}
\usepackage{physics}
\usepackage{gensymb}
\usepackage{dsfont}
\usepackage{epigraph}
\usepackage{dirtytalk}
\usepackage{url}
\usepackage[colorinlistoftodos]{todonotes}
\usepackage{hyperref}
\usepackage{tikz}
\numberwithin{equation}{section}

\title{\boldmath Entangled Universes/Eternal Black Holes Correspondence?}

\author[a,1]{Walid Alhajj,\note{This work is done in part at the Institut de Mathématiques de Bourgogne, Université de Bourgogne Franche-Comté during our master's research }}
\author[a,1]{S. Morteza Hosseini}

\affiliation[a]{Institut de Mathématiques de Bourgogne, Université de Bourgogne Franche-Comté,\\9 Avenue Alain Savary, Dijon, France}

\emailAdd{walidalhajj1999@gmail.com}
\emailAdd{smhroot@gmail.com}

\abstract{In this short paper we conjecture a correspondence between the creation of the universe in entangled pairs and eternal black holes. We shall see that this correspondence will restore the matter-antimatter asymmetry at the beginning of the universe and provide a new cosmological model that can be used to map the physics of the entire universe to the physics of the black hole horizon.}  


\begin{document} 
\maketitle
\flushbottom
\section{Introduction}
\label{sec:intro}
The gauge/gravity duality is one of the exciting theories that relate to two pillars of twentieth-century physics, the gravity theory, and the quantum field theory. This duality states that any gravity theory in a given spacetime can be described by a conformal field theory on the boundary of this spacetime. We will focus on one of these dualities which is the AdS/CFT correspondence that relates the gravity theories living in the anti-de Sitter spacetime to the conformal field theories on its boundary. It was first noticed in 1974 by the Gerard 't Hooft \cite{tHooft:1973alw} that there is a duality between the string theory and quantum field theory calculations of QCD theories with color charges. Then he followed this paper with another paper in 1993 \cite{tHooft:1993dmi} that showed that our world on a Planck scale is not 3+1 dimensional but rather it can be defined on a two-dimensional lattice evolving with time. Furthermore, in 1995 Leonard Susskind published a paper \cite{Susskind:1994vu} which states that the information of n+1 dimensional spacetime is encoded on the boundary of this spacetime. In 1997, this duality becomes clear by Maldacena's conjecture of the AdS/CFT correspondence \cite{Maldacena:1997re} by looking at the different perspectives of the D-branes in supergravity. This duality gives new advances to our understanding of quantum gravity. The beauty of such a duality appears when we turn out to the computations. For example, calculating the entropy of the strongly coupled field theory is very hard. With the help of the correspondence, this problem transforms into the problem of computing the entropy of the black hole which is easy to handle with it \cite{Mukherji:2002de}. Regardless that there is no formal mathematical proof of this conjecture, it made a tremendous success in many areas of theoretical physics and it has undergone many successful checks in the last decades.

	Baryonic asymmetry is one of the unsolved puzzles in the modern cosmological model \cite{Dine:2003ax,Morrissey:2012db,2009GReGr..41.1455M}. Even though this asymmetry has been observed throughout the universe in different redshifts consistently, there has been no evidence for the baryonic matter being more probable to form than antimatter. The exact mechanism under which such asymmetry could arise has not been found but the general idea is that one of the following three Sarkhof's conditions \cite{Sakharov:1967dj} must be satisfied to have an early unbalance between the amount of matter and antimatter,
	\begin{itemize}
		\item Breaking B-symmetry
		\item C-asymmetry and violation of CP-invariance 
		\item Thermal in-equilibrium.
	\end{itemize} 

In recent years, however, there have been few studies on the possible solutions of the Friedmann equation that proposed the universe has been formed along with another universe as a universe-antiuniverse pair \cite{Robles-Perez:2021rqt,Robles-Perez:2020fnn}. This work is a short discussion on the universe-antiuniverse pair in the AdS/CFT framework and here we will propose a correspondence between the entangled universe-antiuniverse pair and the eternal black hole. 
The ideas presented here are going to be based on the AdS/CFT framework and some recent developments in it. Since observational data suggest that dS spacetime is more suitable to describe our universe, we will generally consider AdS$_5$ spacetime with a 4-dimensional dS spacetime as its boundary.

\section{Entangled Universes/Eternal Black Holes Correspondence} 
In the past decade, there have been numerous attempts to study quantum entanglement in different frameworks such as AdS/CFT. Many studies are towards understanding how classical geometry and distance could emerge from this very specific behavior of the quantum system, entanglement. For example, the author in \cite{VanRaamsdonk:2010pw} has essentially argued that using holographic entanglement entropy one can realize that as two quantum systems are more and more entangled and their entanglement entropy increases, they will have less physical distance and get closer to each other on the AdS side. This paper has become one of the fundamental pillars of the ongoing research of building spacetime using quantum entanglement and one of its measures, entanglement entropy. In this section, we will conjecture a correspondence between entangled universes and eternal black holes.    
	
The state of entanglement between two universes can be studied in more detail through different stages of the universe with observations and experiments. Our general belief is that if such entangled universes exist then they should either be less entangled or completely disentangled very rapidly. Temperature arguments presented both in \cite{Maldacena:2001kr} and \cite{VanRaamsdonk:2010pw} also give us more reason to believe that the entanglement between two universes was stronger in the beginning as both were too hot and as time passes by they slowly got colder and naturally got further away from each other to the point where this physical connection could be well out of cosmic horizon for observers in each one of the universes. 

 Let us propose that our universe was created in disconnected pairs. In the viewpoint of quantum entanglement, the Hilbert space of the universe will be given as a tensor product of the Hilbert spaces of the two disconnected regions,
\begin{equation}
\mathcal{H}=\mathcal{H}_{1}\otimes\mathcal{H}_{2}
\end{equation}
where $\mathcal{H}_{1}$ and $\mathcal{H}_{2}$ are Hilbert spaces of both subsystems respectively. 
Then, the quantum state of the universe is then given by,
\begin{equation}\label{5}
\ket{\Psi}=\ket{\Psi_{1}}\otimes\ket{\Psi_{2}}
\end{equation}

Since the two regions are disconnected, their CFTs do not interact in any way. We will consider now a state in which the two subsystems entangled \cite{VanRaamsdonk:2010pw},
\begin{equation}\label{1}
\ket{\Psi}=\sum_{n}e^{-\frac{\beta E_{n}}{2}}\ket{n}\otimes\ket{n}
\end{equation}
where $\ket{n}$ is the number eigenstate of a single CFT in $S^{3}$. Thus, the state $\ket{\Psi}$ is the quantum superposition of the two disconnected spacetime regions. However, we can notice that $\ket{\Psi}$ also correspondence to the state of eternal black hole spacetime \cite{Maldacena:2001kr} which represents two maximally entangled black holes in disconnected spacetime regions at time t=0. Therefore, we conjecture a correspondence between the pair creation of the universe and the eternal black holes.

\section{Implications}
We will examine some of the interesting features that emerge from this correspondence. From \cite{Maldacena:2013xja}, the time evolution of the state \eqref{1} can be written in the form, 
\begin{equation}\label{2}
\ket{\Psi(t)}=\sum_{n}e^{-\frac{\beta E_{n}}{2}}e^{-2iE_{n}t}\ket{n}\otimes\ket{\bar{n}}
\end{equation}
where $\ket{\bar{n}}$ is the CP conjugate of the state $\ket{n}$. By the correspondence, we expect that the time evolution of the pair creation of the universe is governed by \eqref{2}. The appearance of the conjugate number eigenstate $\ket{\bar{n}}$ on the cosmological side implies that one of the subsystems is filled with matter while the other is filled with antimatter. So, we end up with the creation of universe-antiuniverse pair which restores the matter-antimatter asymmetry from the perspective of a single universe. 

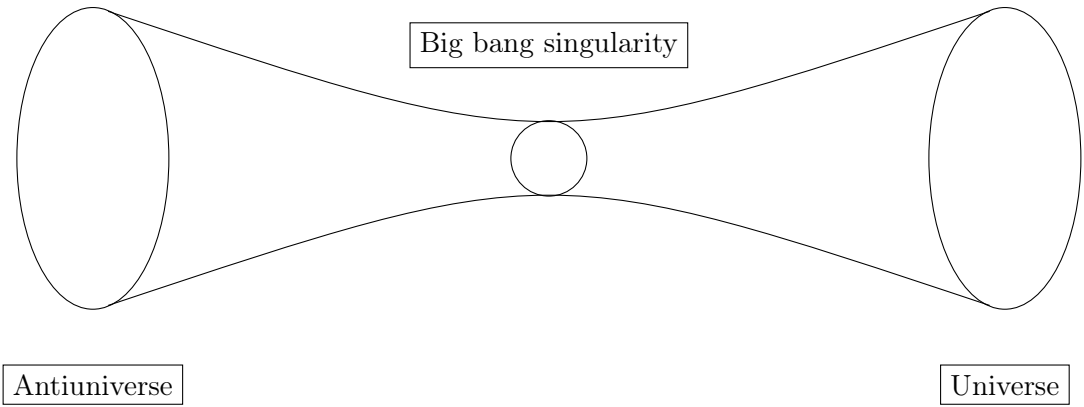
\begin{figure}
\centering
\begin{tikzpicture}
\draw (2,2) ellipse (1cm and 2cm);
\draw (14,2) ellipse (1cm and 2cm);
\draw (2.2,3.95) .. controls (8,2)  .. (13.8,3.95);
\draw (2.2,0.05) .. controls (8,2)  .. (13.8,0.05);
\draw (8,2) circle (0.5cm);

\node[draw] at (2,-1) {Antiuniverse};
\node[draw] at (14,-1) {Universe};
\node[draw] at (8,3.5) {Big bang singularity};
\end{tikzpicture}
\caption{2D schema of an entangled Universe-Antiuniverse pair connected by ER bridge with the big bang singularity as a throat of that bridge}
\label{ua}
\end{figure}

Another interesting feature that emerges from the entangled universes/eternal black holes is that it gives a new vision of our universe. By ER=EPR conjecture \cite{Maldacena:2013xja}, the geometries of the two entangled black holes are connected by some kind of complex Einstein-Rosen bridge \cite{PhysRev.48.73}. We propose, by the correspondence, that the two branches of the universe are also connected by some kind of higher dimensional ER bridge. The throat of this ER bridge can be represented by a hypersurface of the minimal area of the universe's beginning. Imagine that we live in one of these branches, then our universe, by this proposition, is just a boundary of this bridge. If the conjecture is true, it seems that our universe at this moment appears like some kind of cosmic horizon similar to a black hole horizon.

Let us make a thought experiment of an observer living inside that horizon. If this observer emits a light beam now, he can never catch that beam in the future even after an infinity of years, unless he moves faster than the speed of light. We can argue from that thought experiment that the cosmic horizon encodes inside it every moment in the history of the universe. Also, it can be noticed that the horizon behaves like a firewall that prevents any outside observer to enter the horizon.  

We expect, by the correspondence, that the cosmic horizon should also radiate, and hence it has a thermodynamic interpretation. In contradiction with the black hole horizon, our cosmic horizon should radiate inward. It has been known that the entropy of the black hole is finite and is given by the Bekenstein-Hawking formula \cite{PhysRevD.13.191},
\begin{equation}\label{b}
S=\frac{A}{4G}
\end{equation}
where $A$ is the surface area of the event horizon of the black hole. We assume, due to the correspondence, that the entropy of the cosmic horizon is also governed by the Bekenstein-Hawking formula \eqref{b} where $A$ should present the area of the cosmic horizon. Thus, we could have a better understanding of the initial low entropy of the universe \cite{Lineweaver2014} and why the entropy of the universe is always increasing from such a thermodynamic relation.

\section{Conclusion}
The proposed correspondence between an eternal black hole and the entangled pair of universe-antiuniverse can provide a new cosmological model in which the physics of the universe becomes the physics of the black hole horizon. The main elements of this correspondence are the singularity and thermodynamics of each system. This proposition needs to be studied in many different mathematical and physical aspects. Such correspondence will provide the possibility of studying the grand scheme of the evolution of the universe by the evolution of the black hole horizon. Another valuable result of this correspondence is the recovered baryonic symmetry in the form of two entangled universes, one with more matter and another with more antimatter. Our hope is that through this conjecture we can ask more fundamental questions about different current unsolved topics in modern cosmology such as dark matter and dark energy and study them using the different known aspects of black hole horizon and thermodynamics in future work. 

\section{Acknowledgements}
We would like to thank Taro kimura for his comments and advice on this work. WA was partly supported by EIPHI Graduate School (No. ANR-17-EURE-0002), and the Bourgogne-Franche-Comté region.    




\small
\nocite{*}
\bibliographystyle{JHEP}
\bibliography{refs}
 


\end{document}